\documentstyle[12pt,amssymb,epsf,amsbsy]{article}

\newcommand{\bs}[1]{\boldsymbol{#1}}
\def\dfrac#1#2{{\displaystyle {#1 \over #2}}}%
\begin{document}

\title{On the Spinning Motion of the Hovering Magnetic Top.}
\author{P. Flanders \\
The Department of Physics and Astronomy,\\
University of Pennsylvania,\\
PA 19104-6396, USA \and S. Gov S. Shtrikman\thanks{%
Also with the Department of Physics, University of California, San Diego, La
Jolla, 92093 CA, USA} \\
The Department of Electronics,\\
Weizmann Institute of Science,\\
Rehovot 76100, Israel \and H. Thomas \\
The Department of Physics and Astronomy\\
University of Basel\\
CH-4056 Basel, Switzerland}
\maketitle

\begin{abstract}
In this paper we analyze the spinning motion of the hovering magnetic top.
We have observed that its motion looks different from that of a classical
top. A classical top rotates about its own axis which precesses around a
vertical fixed external axis. The hovering magnetic top, on the other hand,
has its axis slightly tilted and moves {\em rigidly} as a whole about the
vertical axis. We call this motion synchronous, because in a stroboscopic
experiment we see that a point at the rim of the top moves synchronously
with the top axis.

We show that the synchronous motion may be attributed to a small deviation
of the magnetic moment from the symmetry axis of the top. We show that as a
consequence, the minimum angular velocity required for stability is given by 
$\sqrt{4\mu HI_{1}/I_{3}^{2}}$ for $I_{3}>I_{1}$ and by $\sqrt{\mu
H/(I_{3}-I_{1})}$ for $I_{3}<I_{1}$. Here, $I_{3}$ and $I_{1}$ are the
principal and secondary moments of inertia, $\mu $ is the magnetic moment,
and $H$ is the magnetic field. For comparison, the minimum angular for a
classical top is given by $\sqrt{4\mu HI_{1}/I_{3}^{2}}$ both for $%
I_{3}<I_{1}$ and for $I_{3}>I_{1}$.

We also give experimental results that were taken with a top whose moment of
inertia $I_{1}$ can be changed. These results show very good agreement with
our calculations.
\end{abstract}

\section{Introduction.}

\subsection{The hovering magnetic top.}

The hovering magnetic top is an ingenious device that hovers in mid-air
while spinning. It is marketed as a kit in the U.S.A. and Europe under the
trade name Levitron$^{TM}$ \cite{levitron,patent} and in Japan under
the trade name U-CAS\cite{ucas}. The whole kit consists of three main parts:
A magnetized top which weighs about 18gr, a thin (lifting) plastic plate and
a magnetized square base plate (base). To operate the top one should set it
spinning on the plastic plate that covers the base. The plastic plate is
then raised slowly with the top until a point is reached in which the top
leaves the plate and spins in mid-air above the base for about 2min. The
hovering height of the top is approximately 3.0 cm above the surface of the
base whose dimensions are about 10cm$\times $10cm$\times $2cm. The kit comes
with extra brass and plastic fine tuning weights as the apparatus is very
sensitive to the weight of the top. It also comes with two wedges to balance
the base horizontally.

The hovering magnetic top serves as an excellent model to study many
branches of physics. These include magnetism, electricity, hydrodynamics,
mechanics of rigid body, wave mechanics, dynamical stability of non linear
as well as linear systems, life time of quasi-stationary states and solid
state magnetism to name just a few. For example, in the field of neutral
particle trapping the hovering magnetic top is a vivid macroscopic device
which demonstrates how it is possible to trap a neutral particle with spin
in a static magnetic field. In fact, one of the recent examples of such
traps was devised and applied to trap Rb$^{87}$ atoms and to form
Bose-Einstein condensation\cite{bec,bec2}. The trap used in this experiment
is based on a rotating magnetic field and was given the name TOP
(Time-averaged Orbiting Potential) trap. Though the TOP trap consists of a
time dependent field it nevertheless operates on roughly the same principles
as the trap of the hovering magnetic top. In addition to understanding
magnetic traps there are many other issues that can be addressed by studying
the hovering magnetic top. Questions such as: How high can it hover?, How
long does it hover?, What are the tolerances on the mass of the top and the
tilt angle of the base? How the trap works in the quantum regime? and other
questions have been studied recently \cite
{time,power1,high,power2,tolerance,quantum} whereas other questions such as
how does friction affect the stability of the top are currently under study.

\subsection{The synchronous motion.}

The physical principles underlying the dynamical stability of the hovering
magnetic top rely on the so-called `adiabatic approximation' and have
recently been discussed in several papers \cite{Berry,Simon,dynamic,Berry2}.
In these articles the top was modeled as an axially symmetric shape with its
magnetization taken as a dipole pointing along the symmetry axis and
situated at the center of mass. Observing the top as it hovers forced us
however, to augment this picture by assuming that the dipole is canted by a
small angle $\Delta $ with respect to the symmetry axis of the top. We were
motivated by the fact that the motion of the magnetic top as it hovers
looked different from that of a classical top (but also by recalling that it
would be forced by manufacturing tolerances). As a classical top slows down
it starts to deviate from the vertical while it precesses with a frequency
which is not synchronous to the frequency of the spin of the top around its
axis. The axis of the hovering magnetic top is also canted to the vertical
but the motion is synchronous. There is only one frequency involved. This
can even be seen with the naked eye. The axis of the magnetic top rotates
around the vertical but {\em synchronously} with the spin of the top around
itself. An analogy, though not literal, is helpful here: The motion of the
hovering magnetic top is reminiscent of the motion of the moon around the
earth but that of a classical top is more like the rotation of the earth
around the sun. The astute observer watching the hovering magnetic top will
note that the canting increases as the top slows down, even with the naked
eye. It was more conspicuous to us when we observed it with a stroboscope
preferably strobed at twice or three times the synchronous frequency. Using
the latter frequency enabled also to measure the phase of the tilt. It is
this way that we found this phase is correlated to the phase of $\Delta $-
the canting of the magnetic moment with respect to the axis of the top. We
reach the same conclusion by observing in slow motion videos of hovering
tops in which we changed $\Delta $ artificially. As will be shown in the
following the canting $\Delta $ explains all the above observations.

The present paper is in fact a study of the effect of this small canting $%
\Delta $ on the hovering of the magnetic top. In addition to explaining the
above qualitative features we also derive quantitative results. In
particular we show that the minimum angular velocity required for stability $%
\omega _{min}$ is given by 
\[
\omega _{min}=\left\{ 
\begin{array}{ll}
\sqrt{\dfrac{4\mu HI_{1}}{I_{3}^{2}}} & \mbox{ for }I_{3}<I_{1} \\ 
\sqrt{\dfrac{\mu H}{I_{3}-I_{1}}} & \mbox{ for }I_{3}>I_{1}.
\end{array}
\right. 
\]
Here, $I_{3}$ and $I_{1}$ are the principal and secondary moment of inertia, 
$\mu $ is the magnetic moment of the top and $H$ is the magnetic field. This
result is to be contrasted with the minimum speed of a classical top\cite
{landau} which is given by $\sqrt{4\mu HI_{1}/I_{3}^{2}}$ for both $%
I_{3}<I_{1}$ and $I_{3}>I_{1}$.

On the experimental side, we have measured the minimum speed of a hovering
top whose moment of inertia $I_{1}$ can be changed so as to cover values
both below and above $I_{3}$. These results show excellent agreement with
our calculations.

\subsection{The structure of this paper.}

The structure of this paper is as follows: In Sec.(\ref{sec2}) we carry out
a detailed analysis of the synchronous motion. We first describe precisely
what is meant by the synchronous motion and define our notations in Sec.(\ref
{sec2.1}). In Sec.(\ref{sec2.2}) we study the conditions that must be met in
order for the system to reach equilibrium. Next, in Sec.(\ref{sec2.3}) we
carry out a dynamical stability analysis to find under what conditions the
equilibrium solution found previously is indeed stable. Finally, in Sec.(\ref
{sec2.4}) we combine all the conditions that were found both from
equilibrium considerations and dynamical stability considerations, and
arrive at the expressions for the minimum angular velocity. The description
of our experiment and a comparison of the results with the theoretical
calculations from the previous section is given in Sec.(\ref{sec3}). In Sec.(%
\ref{sec4}), we summarize our results and discuss possible extensions of the
calculations presented in this paper and other related subjects.

\section{\label{sec2}Analysis of the Synchronous motion.}

\subsection{\label{sec2.1}Description of the Problem.}

\begin{figure}[here]
\begin{tabbing}
\epsfxsize=3.8233in
\epsffile{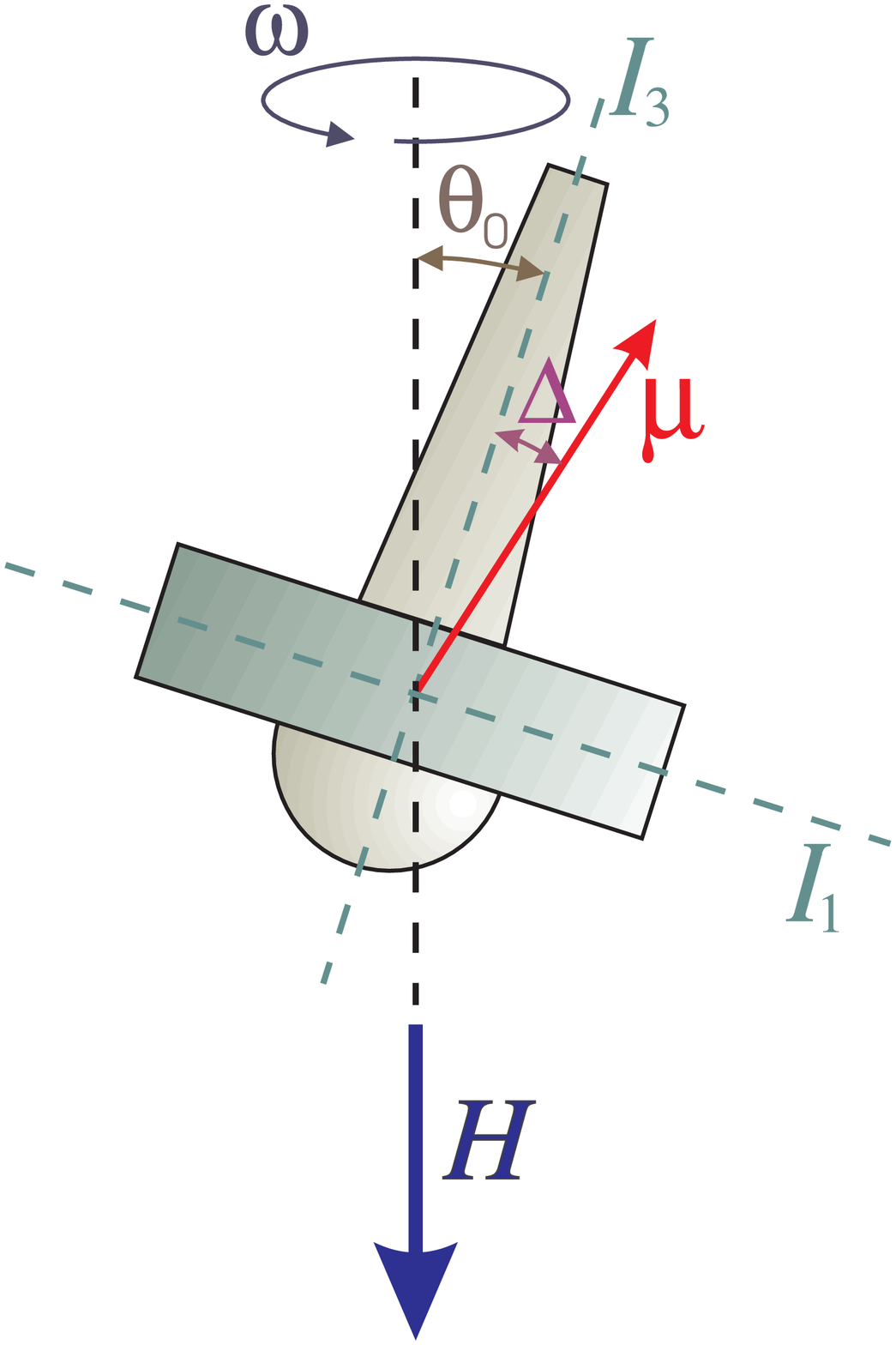}
\end{tabbing}
\caption{The synchronous motion:
The top revolves rigidly around the vertical $\hat{\bs{z}}$ axis with angular
velocity $\omega $. The principal axis of the top (axis $\hat{\bs{n}}$)
makes an angle $\theta _{0}$ with the vertical. The whole system is in a
uniform magnetic field, $H$, pointing downward. The top possesses a magnetic
moment dipole, $\mu $, that makes an angle $\Delta $ with its principal axis.}
\label{fig1}
\end{figure}

As the top hovers above the magnetized plate, the magnetic lift force
exerted by the base is balanced by the top weight. Therefore, to simplify
further calculations we assume that gravity is zero and the magnetic field
is homogeneous. Further, we disregard the translational motion of the top
and assume that it move around a fixed point. This point is taken as the
common origin of the moving and fixed system of coordinates. With these
simplifications the synchronous motion can be modeled as shown in Fig.(\ref
{fig1}).

The figure describes a top that rotates rigidly around
the vertical $\hat{\bs{z}}$
axis with angular velocity $\omega $. The principal axis of the top (axis
 $\hat{\bs{n}}$) makes an angle $\theta _{0}$ with the vertical axis. The
whole system is in a uniform magnetic field $H$, pointing downward. The top
possesses a magnetic moment dipole, $\mu $, that makes an angle $\Delta $
with its principal axis.

\subsection{\label{sec2.2}Equilibrium considerations.}

We first determine the angle between the top axis $\hat{\bs{n}}$ and the
vertical axis $\hat{\bs{z}}$ from torque equilibrium. The angular momentum
of a symmetric top has the form\cite{milne} 
\begin{equation}
\bs{L}=I_{3}\omega _{3} \hat{\bs{n}}+I_{1} \hat{\bs{n}}\times \dfrac{d \hat{\bs{n}}}{dt}  \label{eq11}
\end{equation}
where $\omega _{3}$ is the component of the vector angular
 velocity $ \mbox{$\hat \omega$}$ in the direction of the $\hat{\bs{n}}$ axis.
Since $\hat{\bs{n}}$ is fixed in the body, 
\begin{equation}
\dfrac{d\hat{\bs{n}}}{dt}=\mbox{\boldmath $\omega$} \times \hat{\bs{n}},  \label{eq12}
\end{equation}
and therefore 
\begin{equation}
\bs{L}=\left( I_{3}-I_{1}\right) \omega _{3}\hat{\bs{n}}+I_{1} \mbox{\boldmath $\omega$} .
 \label{eq13}
\end{equation}
For the synchronous motion, the gyroscopic torque in the co-moving frame
must compensate the magnetic torque, hence 
\begin{equation}
\mbox{\boldmath $\omega$} \times \bs{L}=\mbox{\boldmath $\mu$} \times \bs{H}.  \label{eq13.1}
\end{equation}
Thus, $\mbox{\boldmath $\mu$}$ has to be coplanar with $\hat{\bs{z}}$ and $\hat{\bs{n}}
$, and one obtain, with $\omega _{3}=\omega \cos \theta _{0}$, the
equilibrium condition

\begin{equation}
(I_{3}-I_{1})\omega ^{2}\sin \theta _{0}\cos \theta _{0}=\mu H\sin (\theta
_{0}+\Delta ).  \label{eq14}
\end{equation}

We assume that the angle $\Delta $ is small. Then, during its motion the top
axis $\hat{\bs{n}}$ stays close to the $\hat{\bs{z}}$ axis, and we may use
small angle approximation in Eq.(\ref{eq14}) and express $\theta _{0}$ in
terms of $\omega $ as follows: 
\begin{equation}
\theta _{0}\simeq \frac{\mu H\Delta }{(I_{3}-I_{1})\omega ^{2}-\mu H}.
\label{eq20}
\end{equation}
In order to understand what Eq.(\ref{eq20}) means we must resort to a more
realistic description of the hovering top. If we take into account air
friction and the inhomogeneity of the magnetic field, the following picture
emerges: As the top hovers, it experience air friction which causes $\omega $
to decrease with time. Thus, according to Eq.(\ref{eq20}) $\theta _{0}$
increases provided that $I_{3}>I_{1}$ (a disk-like top). This, on the other
hand, decreases the repulsive force on the top which is caused by the field
gradient. Thus, the top rebalances itself at a lower height. This process
continues until the field gradient reaches its maximum value, that occurs at
a height equal roughly to half the size of the base. At this point the
magnetic field can no longer support the top and the latter falls down
vertically. In our experiments we could clearly see that $\theta _{0}$
increases gradually by the naked eye. We conclude that when $\theta _{0}$ is
`large enough' the top will fall down. We denote by $\theta _{0}^{*}$ the
maximum allowed angle. The angular velocity which correspond to this angle
is, according to Eq.(\ref{eq20}), given by 
\begin{equation}
\omega =\sqrt{\dfrac{\mu H\left( \dfrac{\Delta }{\theta _{0}^{*}}+1\right) }{%
I_{3}-I_{1}}}.  \label{eq20.0}
\end{equation}
To estimate $\theta _{0}^{*}$ we recall that the tolernace on the weight is
equal to the toleance allowed on the lift. It follows that\cite{tolerance} 
\[
\dfrac{dm}{m}\sim \frac{\left( \theta _{0}^{*}\right) ^{2}}{2}. 
\]
Since the tolerance on the weight is about $2\%$ (see Ref.\cite{tolerance})
it follows that $\theta _{0}^{*}\sim 0.1$rad. Although we did not measure $%
\Delta $ directly, our experiments, including those in which we have
artificially changed $\Delta $ indicate that $\Delta /\theta _{0}^{*}\ll 1$.
It is therefore justified to neglect this term in Eq.(\ref{eq20.0}) and we
arrive to an approximate expression for the minimum speed when $I_{3}>I_{1}$%
, namely: 
\[
\omega_{min}^{stat}\simeq \sqrt{\dfrac{\mu H}{I_{3}-I_{1}}}%
\mbox{ for }I_{3}>I_{1}. 
\]

For a rod-like top ($I_{3}<I_{1}$) the synchronous motion is also possible,
but now $\Delta $ must be more negative than $-\theta _{0}$. This is because
in this case the direction of the centrifugal torque is {\em reversed, }and
for equilibrium to exist the magnetic torque must also be reversed. The
condition that $\Delta <-\theta _{0}$ gives $\omega >0$ which signifies that
as far as the equilibrium condition is concerned there is no minimum speed
for this case. Following similar arguments as before it can be easily
deduced that in this case $\theta _{0}$ {\em decreases} with time and the
top slowly {\em gains} height as it slows down due to air friction. This is
opposite of what we have found for a disk-like top.

Note also that the rod-like top exhibits {\em diamagneic}-like behavior.
Namely, the differential susceptibility is in this case {\em negative} as
can be seen from Eq.(\ref{eq20}).

Summarizing our results so far we conclude that as far as the equilibrium
condition is concerned the minimum allowed angular velocity depends on the
eccentricity of the top and is given by

\begin{equation}
\omega _{min}^{stat}=\left\{ 
\begin{array}{cc}
\sqrt{\dfrac{\mu H}{I_{3}-I_{1}}} & \mbox{ for a disk-like top } (I_{3}>I_{1}) \\ 
0 & \mbox{ for a rod-like top } (I_{3}<I_{1}) \mbox{,}
\end{array}
\right.
\label{eq20.1}
\end{equation}
and is valid both for clockwise and counter-clockwise directions of the
spin. Note that Eq.(\ref{eq20.1}) is not the all answer because the
dynamical stability should also be considered. This is done in the following
section.

\subsection{\label{sec2.3}Dynamic stability considerations.}

\subsubsection{Overview}

Up to this point we have studied conditions that were derived directly from
the equilibrium condition. In this section we study the conditions that must
be met in order to sustain the synchronous motion. We start by reformulating
the equations of motion in terms of the Lagrangian formalism and Euler's
angles, and then add to the synchronous motion a small perturbation. We then
arrive at a set of equations for the perturbational part and find the
conditions that are required for the perturbation to be oscillatory which is
the condition that the unperturbed motion is stable.

\subsubsection{The Lagrangian and the equations of motion.}

\begin{figure}[here]
\begin{tabbing}
\epsfxsize=4.0248in
\epsffile{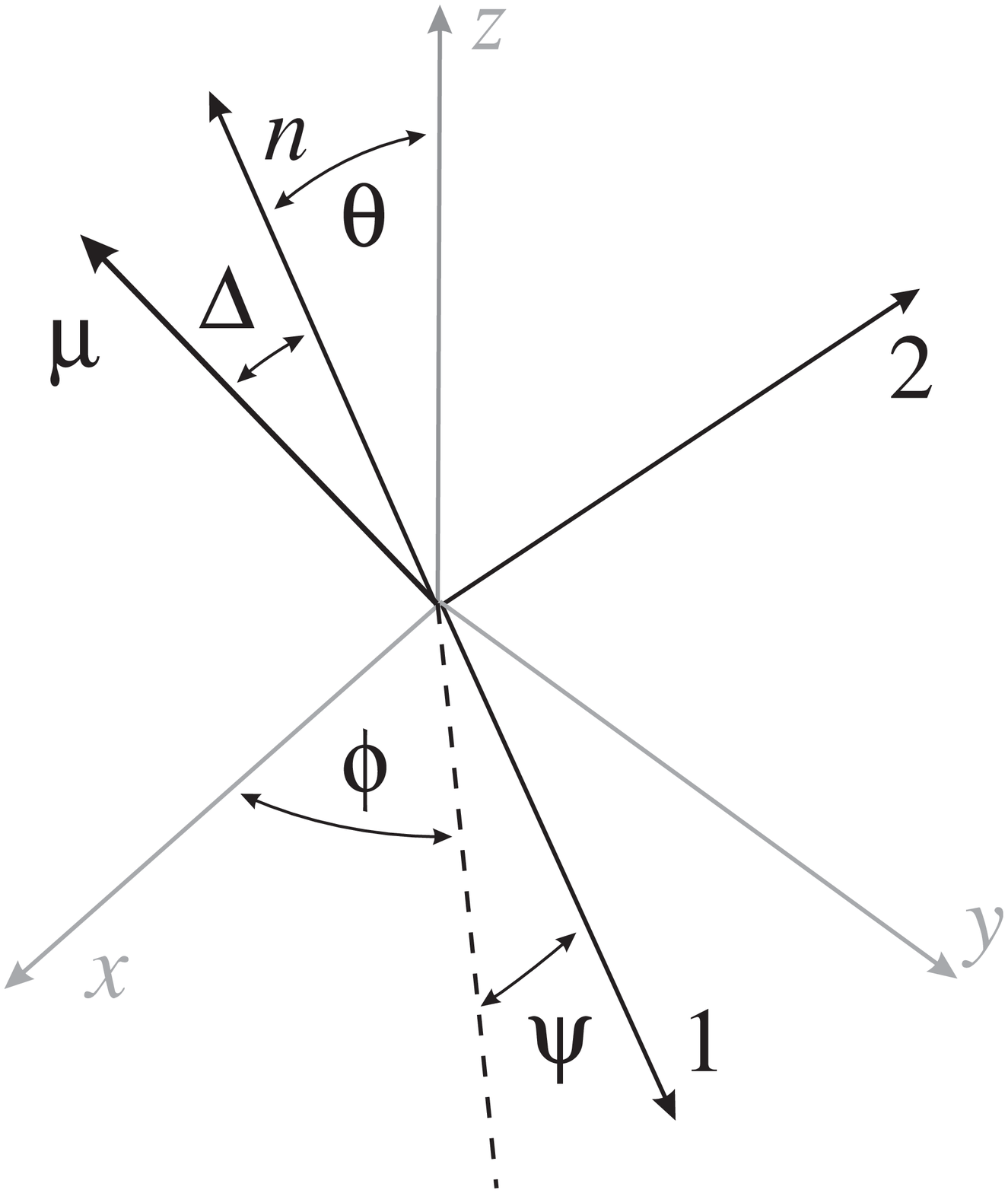}
\end{tabbing}
\caption{Definition of Euler's angles.}
\label{fig2}
\end{figure}

We denote the coordinate system fixed in space
by $(\hat{\bs{x}},\hat{\bs{y}},\hat{\bs{z}})$ and the other that is fixed
in the body by $(\hat{1},\hat{2} ,\hat{\bs{n}})$, as depicted in Fig.(\ref{fig2}).
The magnetic moment $\mbox{\boldmath $\mu$ }$ makes an angle $\Delta $ with the $\hat{\bs{n}}$ axis
and lies in the $\hat{2}$-$\hat{\bs{n}}$ plane. Thus 
\[
\mbox{\boldmath $\mu$ }=\mu \left( \hat{\bs{n}}\cos \Delta -\hat{2}\sin \Delta \right) . 
\]
The magnetic field is uniform and points along the $-\hat{\bs{z}}$
direction, i.e. 
\[
\bs{H}=-H\hat{\bs{z}}=-H\left( \hat{1}\sin \psi \sin \theta +\hat{2}\cos
\psi \sin \theta +\hat{\bs{n}}\cos \theta \right) , 
\]
where $\phi $, $\theta $ and $\psi $ are the first, second and third Euler's
angles, defined in Fig.(\ref{fig2}). Therefore, the magnetic energy for this
configuration has the form 
\[
E_{mag}=-\mbox{\boldmath $\mu$} \cdot \bs{H}=\mu H\left( \cos \theta \cos \Delta -\sin
\Delta \cos \psi \sin \theta \right) . 
\]
The kinetic energy, expressed using Euler's angle, is 
\[
E_{kin}=\dfrac{1}{2}I_{3}\left( \dot{\psi}+\dot{\phi}\cos \theta
\right) ^{2}+\frac{1}{2}I_{1}\left( \dot{\theta}^{2}+\dot{\phi}^{2}\sin
^{2}\theta \right) . 
\]
The Lagrangian $L$ is given by the difference between the kinetic energy $E_{kin}$ and 
the magnetic energy $E_{mag}$, 
\begin{equation}
\begin{array}{c}
L=\dfrac{1}{2}I_{3}\left( \dot{\psi}+\dot{\phi}\cos \theta \right) ^{2}+\dfrac{%
1}{2}I_{1}\left( \dot{\theta}^{2}+\dot{\phi}^{2}\sin ^{2}\theta \right) \\ 
-\mu H\left( \cos \theta \cos \Delta -\sin \Delta \cos \psi \sin \theta
\right) .
\end{array}
\label{eq21}
\end{equation}
The dynamics of motion for this system is then governed by the Lagrange
equations of motion 
\begin{equation}
\begin{array}{c}
\dfrac{d}{dt}\left( \dfrac{\partial L}{\partial \dot{\theta}}\right) -\dfrac{\partial L}{\partial \theta }=0 \\ 
\dfrac{d}{dt}\left( \dfrac{\partial L}{\partial \dot{\psi}}\right) -\dfrac{\partial L}{\partial \psi }=0 \\ 
\dfrac{d}{dt}\left( \dfrac{\partial L}{\partial \dot{\phi}}\right) -\dfrac{\partial L}{\partial \phi }=0.
\end{array}
\label{eq22}
\end{equation}
With the Lagrangian given in Eq.(\ref{eq21}) one obtains

\begin{equation}
\begin{array}{c}
I_{1}\ddot{\theta}+\left[ (I_{3}-I_{1})\dot{\phi}\cos \theta +I_{3}\dot{\psi}%
\right] \dot{\phi}\sin \theta =\mu H\left( \sin \theta \cos \Delta +\sin
\Delta \cos \psi \cos \theta \right) \\ 
I_{3}\dot{\phi}\dot{\theta}\sin \theta -I_{3}\left( \ddot{\psi}+\ddot{\phi}%
\cos \theta \right) =\mu H\sin \Delta \sin \psi \sin \theta \\ 
\left( I_{3}\cos ^{2}\theta +I_{1}\sin ^{2}\theta \right) \dot{\phi}+I_{3}%
\dot{\psi}\cos \theta =L_{z}=\text{Const.}
\end{array}
\label{eq23}
\end{equation}

\subsubsection{The Stationary Solution.}

A stationary solution to the set Eqs.(\ref{eq23}) is found by setting 
\begin{eqnarray}
\theta (t) &=&\theta _{0}=\text{Const.}  \label{eq24.0} \\
\psi (t) &=&\psi _{0}=\mbox{Const.}  \nonumber
\end{eqnarray}
Substituting these into the third equation of the set Eqs.(\ref{eq23}) and
integrating over time yields 
\begin{equation}
\phi (t)=\omega =\mbox{Const.}  \label{eq24.1}
\end{equation}
Using the second equation we find that 
\begin{equation}
\psi _{0}=0,  \label{eq24.2}
\end{equation}
while the remaining (first) equation gives 
\begin{equation}
(I_{3}-I_{1})\omega ^{2}\sin \theta _{0}\cos \theta _{0}=\mu H\sin (\theta
_{0}+\Delta ).  \label{eq25}
\end{equation}
Note that Eq.(\ref{eq25}) is identical to Eq.(\ref{eq14}), which was derived
in Sec.(\ref{sec2.2}).

\subsubsection{Perturbing the stationary solution.}

We now perturb the stationary solution by adding small variations, 
\begin{equation}
\begin{array}{c}
\dot{\phi}(t)=\dot{\phi}_{s}(t)+\delta \dot{\phi}(t)=\omega +\delta \dot{\phi%
}(t) \\ 
\psi (t)=\psi _{s}(t)+\delta \psi (t)=0+\delta \psi (t) \\ 
\theta (t)=\theta _{s}(t)+\delta \theta (t)=\theta _{0}+\delta \theta (t)
\end{array}
\label{eq26}
\end{equation}
Substituting this these into Eqs.(\ref{eq23}) and keeping only first-order
terms yields

\begin{equation}
\begin{array}{c}
\begin{array}{c}
I_{1}\delta \ddot{\theta}+(I_{3}-I_{1})\omega \delta \dot{\phi}\sin (2\theta
_{0}) \\ 
+(I_{3}-I_{1})\omega ^{2}\delta \theta \cos (2\theta _{0})+I_{3}\delta \dot{%
\psi}\omega \sin \theta _{0}=\mu H\delta \theta \cos (\Delta +\theta _{0})
\end{array}
\\ 
I_{3}(\delta \ddot{\psi}+\delta \ddot{\phi}\cos \theta _{0})-I_{3}\delta 
\dot{\theta}\omega \sin \theta _{0}+\mu H\sin \Delta \delta \psi \sin \theta
_{0}=0 \\ 
\begin{array}{c}
-2\omega (I_{3}-I_{1})\delta \theta \sin \theta _{0}\cos \theta
_{0}+I_{3}\delta \dot{\psi}\cos \theta _{0} \\ 
+\delta \dot{\phi}\left( I_{3}\cos ^{2}\theta _{0}+I_{1}\sin ^{2}\theta
_{0}\right) =0
\end{array}
.
\end{array}
\label{eq27}
\end{equation}

Using Eq.(\ref{eq25}) we express $\Delta $ in terms of $\theta _{0}$ and to
simplify further calculations we now restrict ourselves to the case where $%
\theta _{0}$ is a small angle and work to second-order in $\theta _{0}$.
Under the small-angle approximation Eqs.(\ref{eq27}) becomes 
\begin{equation}
\begin{array}{c}
\begin{array}{c}
I_{1}\delta \ddot{\theta}+2(I_{3}-I_{1})\omega \delta \dot{\phi}\theta
_{0}+(I_{3}-I_{1})\omega ^{2}\delta \theta \left( 1-2\theta _{0}^{2}\right)
\\ 
+I_{3}\delta \dot{\psi}\omega \theta _{0}=\mu H\delta \theta \left(
1-0.5\beta ^{2}\theta _{0}^{2}\right)
\end{array}
\\ 
\begin{array}{c}
I_{3}\delta \ddot{\psi}+I_{3}\delta \ddot{\phi}\left( 1-\dfrac{1}{2}\theta
_{0}^{2}\right) -I_{3}\delta \dot{\theta}\omega \theta _{0} \\ 
+(\omega ^{2}(I_{3}-I_{1})-\mu H)\theta _{0}^{2}\delta \psi =0
\end{array}
\\ 
\begin{array}{c}
-2\omega \theta _{0}(I_{3}-I_{1})\delta \dot{\theta}+I_{3}\delta \ddot{\psi}%
\left( 1-\frac{1}{2}\theta _{0}^{2}\right) \\ 
+I_{3}\delta \ddot{\phi}\left( 1-\theta _{0}^{2}\right) +I_{1}\delta \ddot{%
\phi}\theta _{0}^{2}=0.
\end{array}
\end{array}
\label{eq28}
\end{equation}
where $\beta =\left( I_{3}-I_{1}\right) \omega ^{2}/\mu H-2$.

Since Eqs.(\ref{eq28}) are linear in the perturbations and homogenous, the
general solution of these equations is a linear combination of exponential
functions. We therefore set 
\[
\begin{array}{c}
\delta \theta (t)=\delta \theta _{0}e^{\lambda t} \\ 
\delta \phi (t)=\delta \phi _{0}e^{\lambda t} \\ 
\delta \psi (t)=\delta \psi _{0}e^{\lambda t}
\end{array}
\]
and substitute these into Eqs.(\ref{eq28}). The result is the matrix
equation for the determination of the eigenvalues $\lambda $

\[
\bs{A}\left[ 
\begin{array}{c}
\delta \theta _{0} \\ 
\delta \phi _{0} \\ 
\delta \psi _{0}
\end{array}
\right] =\left[ 
\begin{array}{c}
0 \\ 
0 \\ 
0
\end{array}
\right] , 
\]
where $\bs{A}$ is the matrix given by:

\begin{equation}
\left[ 
\begin{array}{ccc}
\left( 
\begin{array}{c}
\lambda ^{2}I_{1}-\mu H(1-0.5\beta ^{2}\theta _{0}^{2}) \\ 
+(I_{3}-I_{1})\omega ^{2}(1-2\theta _{0}^{2})
\end{array}
\right) & 2\lambda (I_{3}-I_{1})\omega \theta _{0} & \lambda I_{3}\omega
\theta _{0} \\ 
-\lambda I_{3}\omega \theta _{0} & \lambda ^{2}I_{3}(1-\dfrac{1}{2}\theta
_{0}^{2}) & \left( 
\begin{array}{c}
\lambda ^{2}I_{3} \\ 
+\left[ \omega ^{2}\left( I_{3}-I_{1}\right) -\mu H\right] \theta _{0}^{2}
\end{array}
\right) \\ 
-2\omega \lambda (I_{3}-I_{1})\theta _{0} & \left( 
\begin{array}{c}
\lambda ^{2}I_{3}(1-\theta _{0}^{2}) \\ 
+\lambda ^{2}I_{1}\theta _{0}^{2}
\end{array}
\right) & \lambda ^{2}I_{3}(1-\frac{1}{2}\theta _{0}^{2})
\end{array}
\right]  \label{eq29}
\end{equation}
For a nontrivial solution to exist, the determinant of the matrix must
vanish. After a few algebraic steps we find

\[
\det \bs{A}=-\lambda ^{2}I_{3}\theta _{0}^{2}\left( a\lambda ^{4}+b\lambda
^{2}+c\right) +{\cal O}\left( \theta _{0}^{4}\right) =0, 
\]
where 
\begin{eqnarray*}
a &\equiv &I_{1}^{2} \\
b &\equiv &-2I_{1}\left[ \mu H-\omega ^{2}(I_{3}-I_{1})\right] -\omega
^{2}(I_{3}-2I_{1})(2I_{1}-I_{3}) \\
c &\equiv &\left[ \mu H-\omega ^{2}(I_{3}-I_{1})\right] ^{2}
\end{eqnarray*}
Thus, we find a doubly degenerate zero eigenvalue corresponding to the
cyclic coordinate $\phi $, which enter the equations of motion only through
its time derivative. The other eigenvalues are the roots of a quadratic
equation in $\lambda ^{2}$.

For the perturbation to be bounded for all times, the non-vanishing
eigenvalues must be purely imaginary. Thus, the two $\lambda ^{2}$ roots
must be both real and negative. The condition for these roots to be real is
that $b^{2}-4ac>0$. This results in the inequality 
\begin{equation}
\dfrac{\mu H}{\omega ^{2}I_{1}}<\frac{1}{4}\left( \dfrac{I_{3}}{I_{1}}\right)
^{2},  \label{eq30}
\end{equation}
which is the well-known stability condition of a classical top with one
point fixed\cite{landau}. For both roots to be negative, their product must
be positive (i.e. $c/a>0$) and their sum must be negative ($-b/a<0$). Since
both $a$ and $c$ are squares of real numbers, their ratio {\em is} positive.
We thus find as the second condition

\begin{equation}
\dfrac{\mu H}{\omega ^{2}I_{1}}<\dfrac{1}{2}\left( \dfrac{I_{3}}{I_{1}}\right)
^{2}-\left( \dfrac{I_{3}}{I_{1}}\right) +1.  \label{eq31}
\end{equation}
It may be easily verified that for a physical top, for which $I_{3}<2I_{1}$,
the inequality Eq.(\ref{eq31}) is satisfied whenever Eq.(\ref{eq30}) holds.
Therefore, the overall condition for the {\em stability} of the synchronous
motion is given by Eq.(\ref{eq30}) alone, namely 
\begin{equation}
\omega _{min}^{dyn}=\sqrt{\dfrac{4\mu HI_{1}}{I_{3}^{2}}}
\label{eq32}
\end{equation}
both for $I_{3}>I_{1}$ and for $I_{3}<I_{1}$.

\subsection{\label{sec2.4}Conclusions.}

Hitherto we have found separate conditions for the minimum angular speed,
one pair given in Eq.(\ref{eq20.1}) arising from equilibrium considerations,
the other one given in Eq.(\ref{eq32}) originating from dynamical stability
considerations. Since these conditions must be satisfied simultaneously, we
now take the {\em union} of these conditions. Taking into account the fact
that for a disk-like top $\sqrt{\mu H/(I_{3}-I_{1})}$ is larger than $\sqrt{%
4\mu HI_{1}/I_{3}^{2}}$, we find that the union of these conditions amounts
to the following expressions for $\omega _{min}$: 
\begin{equation}
\omega _{min}=\left\{ 
\begin{array}{cc}
\sqrt{\dfrac{\mu H}{I_{3}-I_{1}}} & \mbox{ for a disk-like top } (I_{3}>I_{1}) \\ 
\sqrt{\dfrac{4\mu HI_{1}}{I_{3}^{2}}} & \mbox{ for a rod-like top } (I_{3}<I_{1}).
\end{array}
\right.  \label{eq33}
\end{equation}
According to this result, there should be an abrupt change in $\omega _{min}$ in the vicinity of a sphere-like top ($I_{1}=I_{3}$) even for a
vanishingly small value of the angle $\Delta $. This abrupt change is also
found in our experiments, as is shown in the next section.

\section{\label{sec3}Experimental Results.}

\begin{figure}[here]
\begin{tabbing}
\epsfxsize=4.8732in
\epsffile{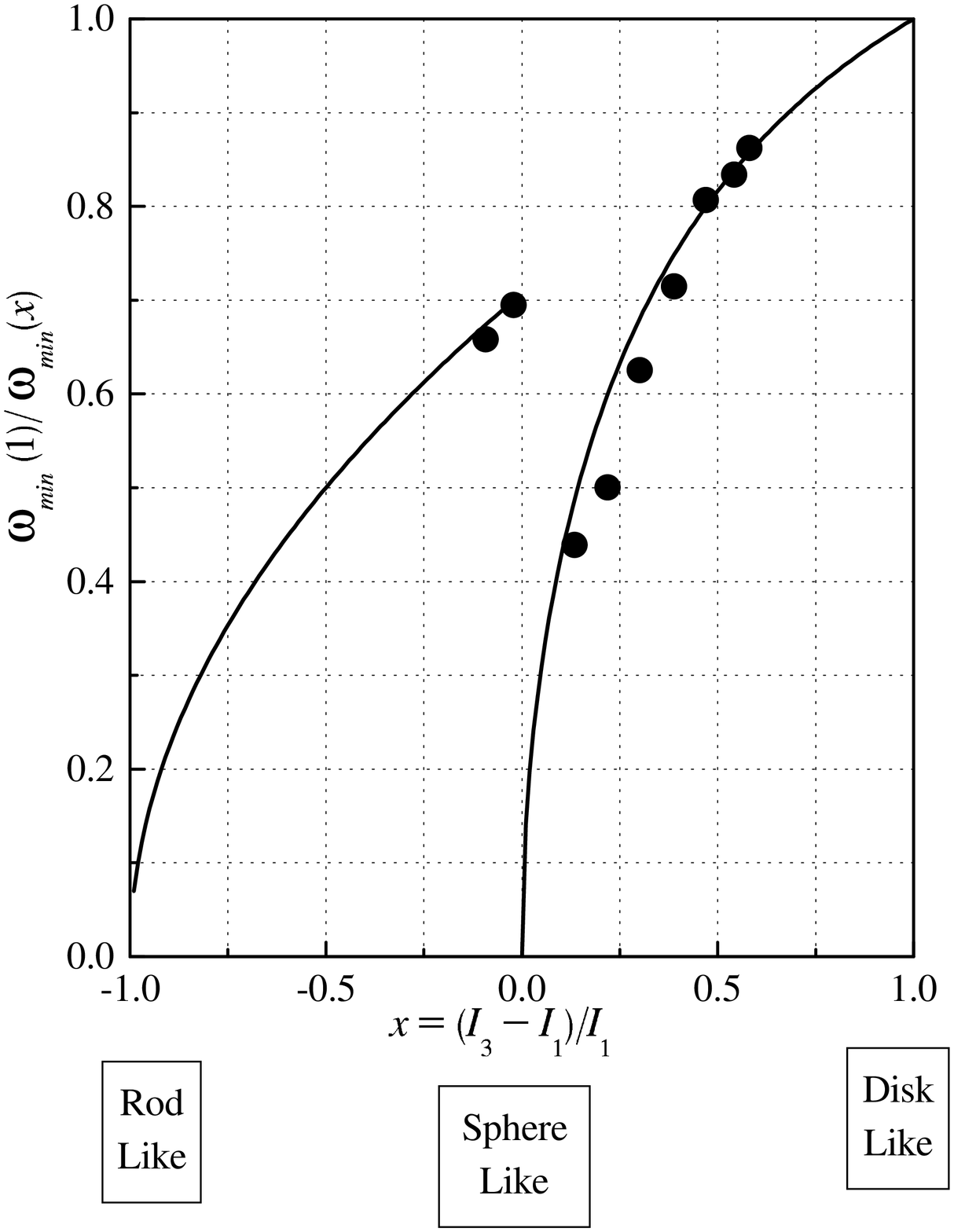}
\end{tabbing}
\caption{Reciprocal of the normalized
minimum angular velocity versus top shape. The bullets are the measured
results and the solid line is the calculated results.}
\label{fig3}
\end{figure}

To test our calculations we have built a magnetized top whose moment of
inertia $I_{1}$ can be changed so as to cover values both below and above $%
I_{3}$. We have measured the minimum angular speed of the top for several
values of $I_{1}/I_{3}$. Fig.(\ref{fig3}) shows (in bullets) the reciprocal
of the measured angular speed normalized to the minimum angular speed of a
pure disk top versus $x\equiv (I_{3}-I_{1})/I_{1}$. To compare these results
to our calculation we rewrite Eq.(\ref{eq33}) as an equation for the
normalized reciprocal angular speed $\omega _{min}\left( 1\right)
/\omega _{min}\left( x\right) $ in terms of $x$. The result is 
\[
\dfrac{\omega _{min}\left( 1\right) }{\omega _{min}\left(
x\right) }=\left\{ 
\begin{array}{cc}
\sqrt{\dfrac{2x}{x+1}} & ;\mbox{ }0<x<1 \\ 
\sqrt{\dfrac{x+1}{2}} & \mbox{ };-1<x<0.
\end{array}
\right. 
\]
Fig.(\ref{fig3}) shows $\omega _{min}\left( 1\right) /\omega _{min}\left( x\right) $ as a solid line. It is clear from the figure that the
experimental results support our theory.

The two main features of our theoretical results are: 1) the divergence of $%
\omega _{min}$ for disk-like tops as they approach spherical shape
and 2) the discontinuity in $\omega _{min}$ as one goes over to
rod-like top. 1) Unfortunately, we could not go beyond $\omega >2.5\omega _{min}
(1)$ because the maximum spin frequency is limited by the dynamic
stability of the top in the trap\cite{dynamic}. It is therefore impossible
to observe experimentally the above divergence. 2) Though we have only two
measured points at the rod-like part of the graph it is conspicuous that
they verify the difference in the origin of the instability for disk- and
rod-like tops that our theory shows.
\section{\label{sec4}Discussion.}

In this paper we have analyzed one possible cause for the synchronous motion
and attributed it to the small difference between the direction of the
principal axis of inertia and the direction of the magnetic moment. We have
shown that as a consequence, for a disk-like top the minimum angular
velocity required for stability {\em differs} from that for a classical top
(with one point fixed) and is given by: 
\[
\omega _{min}=\left\{ 
\begin{array}{ll}
\sqrt{\dfrac{4\mu HI_{1}}{I_{3}^{2}}} & \mbox{ for }I_{3}<I_{1} \\ 
\sqrt{\dfrac{\mu H}{I_{3}-I_{1}}} & \mbox{ for }I_{3}>I_{1}
\end{array}
\right. 
\]
where $I_{3}$ and $I_{1}$ are the principal and secondary moment of inertia, 
$\mu $ is the magnetic moment of the top and $H$ is the magnetic field. This
result is found to be in a very good agreement with our experiments.

We remark that in the calculations presented in this paper we have neglected
some aspects which are relevant to the behavior of the top. For example, the
dipole moment of the top carries with it a minute amount of {\em intrinsic} 
{\em spin} of the order of $S\sim \hbar \mu /\mu _{B}$ where $\hbar $ is
Planck's constant and $\mu _{B}$ is the Bohr's magneton. The existence of
the spin contributes an additional term to the total angular momentum
written in Eq.(\ref{eq12}). In this case the symmetry between clockwise (CW)
and counter-clockwise (CCW) rotations breaks down, and one finds {\em %
different} values for the minimum angular speed for each direction. Due to
the smallness of the spin with respect to the orbital angular momentum, the
CW-CCW asymmetry becomes pronounced only when the {\em size} of the top is
of the order of tenths of microns. Our calculations show that when the sense
of rotation is roughly antiparallel to the direction of the spin the minimum
angular speed increases, whereas when it is roughly parallel to the spin,
the minimum angular speed decreases. This conclusion immediately raises the
question whether it is possible to hover a top using its intrinsic spin
alone without the need to spin it. To study this we assume the top to be
endowed intrincily with spin proportional to its magnetic moment. We find
that it may be possible to hover a top having only this spin with no
additional angular momentum\cite{tobe}.

Another interesting point is what happens if we relinquish the requirement
that $\theta _{0}$ is small. Note that the equilibrium equation Eq.(\ref
{eq14}) is that of the well-known Stoner and Wolfarth model (SW)\cite
{sw1,sw2}, extensively used to described the hysteresis of single domain
ferromagnetic particles. In that case at most only {\em two} of the
equilibria are stable. We find, however, in our case the unstable SW becomes
stable under some conditions, being stabilized by the dynamics. A word of
warning is due here when applying these results to the hovering top. We have
in this work considered only the rotational degrees of freedom, and what is
stable under these is not necessarily stable any more in an inhomogeneous
field necessary to trap the top. We believe however that the small angle
solution is still stable there under the same conditions that are given by us%
\cite{dynamic} for $\Delta =0$.


\begin{thebibliography}{99}
\bibitem{levitron}  The Levitron is available from `Fascinations', 18964 Des
Moines Way South, Seattle, WA 98148.

\bibitem{patent}  Hones et al., U.S. Patent Number: 5,404,062, Date of
Patent: Apr. 4, 1995.

\bibitem{ucas}  The U-CAS is available from Masudaya International Inc.,
6-4, Kuramae, 2-Chome, Taito-Ku, Tokyo, 111 Japan.

\bibitem{bec}  `Physics Today', Aug. 1995, pp. 17-20.

\bibitem{bec2}  W. Petrich, M. H. Anderson, J. R. Ensher and E. A. Cornell, 
{\em Phys. Rev. Lett. } $\bs{74}$, 17, 3352-3355 (1995).

\bibitem{time}  S. Gov, S. Shtrikman and S. Tozik, {\em Ann. Meet. of the
IPS}, $\bs{42}$, 122 (1996).

\bibitem{power1}  S. Gov, H. Matzner and S. Shtrikman, {\em Ann. Meet. of
the IPS}, $\bs{42}$, 121 (1996).

\bibitem{high}  S. Gov and S. Shtrikman, {\em Proc. of the 19}$^{th}$%
{\em \ Israel IEEE Conv.}, 184 (1996).

\bibitem{power2}  S. Gov, H. Matzner and S. Shtrikman, {\em Proc. of the 19}$%
^{th}${\em \ Israel IEEE Conv.}, 121 (1996).

\bibitem{tolerance}  S. Gov, H. Matzner and S. Shtrikman,{\em \ Ann. Meet.
of the IPS}, $\bs{43}$, 47 (1997).

\bibitem{quantum}  S. Gov, S. Shtrikman and H. Thomas, {\em Bulletin of the
Israel Physical Society}, $\bs{44}$, 81 (1998).

\bibitem{Berry}  M. V. Berry, {\em Proc. R. Soc. Lond. A}, $\bs{452}$, 1207
(1996).

\bibitem{Simon}  M. D. Simon, L. O. Heflinger and S. L. Ridgway, {\em Am. J.
Phys. ,} $\bs{65}$ (4), 286 (1997).

\bibitem{dynamic}  S. Gov, S. Shtrikman and H. Thomas, {\em Los-Alamos
E-Print Archive}, http://xxx.lanl.gov/, physics/9803020 (1998).

\bibitem{Berry2}  M. V. Berry, A. K. Geim, {\em IOP Publishing Ltd and The
European Physical Society}, 307 (1997).

\bibitem{landau}  ``Mechanics'' by L. D. Landau and E. M. Lifshitz, {\em %
Pergamon Press}, 3$^{rd}$ Ed., 111-114.

\bibitem{milne}  ``Vectorial Mechanics'' by E. A. Milne, {\em Methuen and
Co. London}, 322-323 (1948).

\bibitem{tobe}  S. Gov, S. Shtrikman and H. Thomas, to be published.

\bibitem{sw1}  E. C. Stoner and E. P. Wohlfarth, {\em Phil. Trans. Roy. Soc. 
}(London) A-240, 599, (1948).

\bibitem{sw2}  see also ``The Physical Principles of Magnetism'' by A. A.
Morish, John Wiley and sons, (1965).
\end{thebibliography}
\end{document}